\documentclass{article}
\usepackage{amssymb, amsmath}
\numberwithin{equation}{section}
\newtheorem{definition}{Definition}[section]
\newtheorem{theorem}[definition]{Theorem}
\newtheorem{proposition}[definition]{Proposition}

\newtheorem{remark}[definition]{Remark}

\def\supp{{\rm supp}}

\begin{document}

\title{Structure of solutions near the initial singularity for the surface-symmetric Einstein-Vlasov system}

\author{Sophonie Blaise Tchapnda \\
Department of Mathematics, Faculty of Science,\\
 University of Yaounde I, PO Box 812, Yaounde, Cameroon\\
\texttt{tchapnda@uycdc.uninet.cm}}
\date{}
\maketitle{}
\begin{abstract}
Results on the behaviour in the past time direction of
cosmological models with collisionless matter and a cosmological
constant $\Lambda$ are presented. It is shown that under the
assumption of non-positive $\Lambda$ and spherical or plane
symmetry the area radius goes to zero at the initial singularity.
Under a smallness assumption on the initial data, these properties
hold in the case of hyperbolic symmetry and negative $\Lambda$ as
well as in the positive $\Lambda$ case. Furthermore in the latter
cases past global existence of spatially homogeneous solutions is
proved for generic initial data. The early-time asymptotics is
shown to be Kasner-like for small data.
\end{abstract}
\section{Introduction}
 Global existence and asymptotics in the future for the surface-symmetric
 Einstein-Vlasov system with cosmological constant have been obtained in
 \cite{tchapnda1}-\cite{tchapnda2}. The present paper deals with the
 analysis in the past time direction.

 In the contracting direction the main result in \cite{rein2} was that solutions of
the surface-symmetric Einstein-Vlasov system with vanishing
cosmological constant exist up to $t=0$ for small initial data,
and then the nature of the initial singularity was analyzed. In
the following these results are generalized to the case with
positive cosmological constant or even negative cosmological
constant and hyperbolic symmetry. Also in the present
investigation we show that these results can be strengthened a lot
for the plane or spherically symmetric case with $\Lambda \le 0$.
We prove in these cases that solutions of the Einstein-Vlasov
system exist on the whole interval $(0,t_0]$ for general initial
data. This is the main result of this paper. An important tool of
the proof is a change of variables inspired by one done by M.
Weaver in \cite{weaver} where she showed existence up to $t=0$ for
a certain class of $T^2$ symmetric solutions of the
Einstein-Vlasov system with vanishing cosmological constant. In
her paper, the general strategy of \cite{isenberg}, using areal
coordinates directly in the contracting direction rather than
conformal coordinates (as in
\cite{andreasson1}-\cite{andreasson3}), is used to so sharpen
global existence results obtained in \cite{andreasson1} and
\cite{andreasson3} for Einstein-Vlasov initial data on $T^3$ with
$T^2$ symmetry. We use the same strategy to sharpen results
previously obtained in \cite{rein2} and \cite{andreasson2}.

Now let us recall the formulation of the Einstein-Vlasov system
which governs the time evolution of a self-gravitating
collisinonless gas in the context of general relativity. All the
particles are assumed to have the same rest mass, normalized to
unity, and to move forward in time so that their number density
$f$ is a non-negative function supported on the mass shell
\[
PM:= \{g_{\alpha\beta}p^{\alpha}p^{\beta} = -1,
 \ p^0 > 0 \},
\]
a submanifold of the tangent bundle $TM$ of the space-time
manifold $M$ with metric $g$ of signature $-+++$. We use
coordinates $(t,x^a)$ with zero shift and corresponding canonical
momenta $p^\alpha$ ; Greek indices always run from $0$ to $3$, and
Latin ones from $1$ to $3$. On the mass shell $PM$ the variable
$p^0$ becomes a function of the remaining variables $(t, x^a,
p^b)$ :
\[
p^0 = \sqrt{-g^{00}}\sqrt{1+g_{ab}p^{a}p^{b}}.
\]
The Einstein-Vlasov system now reads
\begin{align*}
\partial_{t}f + \frac{p^{a}}{p^{0}} \partial_{x^{a}}f -
\frac{1}{p^{0}}\Gamma_{\beta\gamma}^{a} p^{\beta} p^{\gamma}
\partial_{p^{a}}f
= 0 \\ G_{\alpha\beta} + \Lambda g_{\alpha\beta} = 8 \pi T_{\alpha\beta} \\
T_{\alpha\beta} = - \int_{\mathbb{R}^{3}}f p_{\alpha}
p_{\beta}|g|^{1/2} \frac{dp^{1}dp^{2}dp^{3}}{p_{0}},
\end{align*}
where $p_{\alpha} = g_{\alpha\beta} p^{\beta}$,
$\Gamma_{\beta\gamma}^{\alpha}$ are the Christoffel symbols, $|g|$
denotes the determinant of the metric $g$, $G_{\alpha\beta}$ the
Einstein tensor, $\Lambda$ the cosmological constant, and
$T_{\alpha\beta}$ is the energy-momentum tensor.

Here we adopt the definition of spacetimes with surface symmetry,
i.e., spherical, plane or hyperbolic symmetry given in
\cite{rendall1}. We write the system in areal coordinates, i.e.
coordinates are chosen such that $R=t$, where $R$ is the area
radius function on a surface of symmetry. The circumstances under
which coordinates of this type exist are discussed in
\cite{andreasson2} for the Einstein-Vlasov system with vanishing
$\Lambda$, and in \cite{tchapnda2} and \cite{tchapnda3} for the
case with $\Lambda$. In such coordinates the metric takes the form
\begin{equation} \label{eq:1.1}
  ds^2 = -e^{2\mu(t,r)}dt^2 + e^{2\lambda(t,r)}dr^2 + t^2
  (d\theta^2 + \sin_{k}^{2}\theta d\varphi^{2})
\end{equation}
where
\begin{displaymath}
 \sin_{k}\theta := \left\{ \begin{array}{ll}
\sin\theta & \textrm{if $k=1$}\\
1 & \textrm{if $k=0$}\\
\sinh\theta & \textrm{if $k=-1$}
  \end{array} \right.
\end{displaymath}

Here $t > 0$, the functions $\lambda$ and $\mu$ are periodic in
$r$ with period $1$. It has been shown in \cite{rein1} and
\cite{andreasson2} that due to the symmetry $f$ can be written as
a function of
\[
t, r, w := e^{\lambda}p^1 \ \textrm{and} \  F := t^{4}(p^2)^2 +
t^4 \sin_{k}^{2}\theta (p^{3})^{2}, \ \textrm{with} \ r,w \in
\mathbb{R} \ ; \ F \in [0,+\infty[ \ ;
\]
i.e. $f = f(t, r, w, F)$. In these variables we have $p^0 =
e^{-\mu}\sqrt{1 + w^{2} + F/t^{2}}$. After calculating the Vlasov
equation in these variables, the non-trivial components of the
Einstein tensor, and the energy-momentum tensor and denoting by a
dot or by prime the derivation of the metric components with
respect to $t$ or $r$ respectively, the complete Einstein-Vlasov
system reads as follows :
\begin{equation} \label{eq:1.2}
\partial_{t}f + \frac{e^{\mu-\lambda}w}{\sqrt{1+w^{2}+F/t^{2}}}
\partial_{r}f - (\dot{\lambda}w +
e^{\mu-\lambda}\mu'\sqrt{1+w^{2}+F/t^{2}})\partial_{w}f = 0
\end{equation}
\begin{equation} \label{eq:1.3}
e^{-2\mu} (2t\dot{\lambda}+1)+ k - \Lambda t^{2} = 8 \pi t^{2}\rho
\end{equation}
\begin{equation} \label{eq:1.4}
e^{-2\mu} (2t\dot{\mu}-1)- k + \Lambda t^{2} = 8 \pi t^{2}p
\end{equation}
\begin{equation} \label{eq:1.5}
\mu' = -4 \pi t e^{\lambda+\mu}j
\end{equation}
\begin{equation} \label{eq:1.6}
e^{-2\lambda}\left(\mu'' + \mu'(\mu' - \lambda')\right) -
e^{-2\mu}\left(\ddot{\lambda}+(\dot{\lambda}-
\dot{\mu})(\dot{\lambda}+\frac{1}{t})\right) + \Lambda  = 4 \pi q
\end{equation}
where
\begin{equation} \label{eq:1.7}
\rho(t, r) := \frac{\pi}{t^{2}} \int_{-\infty}^{\infty}
\int_{0}^{\infty} \sqrt{1+w^{2}+F/t^{2}} f(t, r, w, F) dF dw =
e^{-2\mu}T_{00}(t, r)
\end{equation}
\begin{equation} \label{eq:1.8}
p(t, r) := \frac{\pi}{t^{2}} \int_{-\infty}^{\infty}
\int_{0}^{\infty} \frac{w^{2}}{\sqrt{1+w^{2}+F/t^{2}}} f(t, r, w,
F) dF dw = e^{-2\lambda}T_{11}(t, r)
\end{equation}
\begin{equation} \label{eq:1.9}
j(t, r) := \frac{\pi}{t^{2}} \int_{-\infty}^{\infty}
\int_{0}^{\infty} w f(t, r, w, F) dF dw = -e^{\lambda +
\mu}T_{01}(t, r)
\end{equation}
\begin{equation} \label{eq:1.10}
q(t, r) := \frac{\pi}{t^{4}} \int_{-\infty}^{\infty}
\int_{0}^{\infty} \frac{F}{\sqrt{1+w^{2}+F/t^{2}}} f(t, r, w, F)
dF dw = \frac{2}{t^{2}}T_{22}(t, r).
\end{equation}

We prescribe initial data at some time $t = t_0 > 0$,
\begin{eqnarray*}
f(t_0, r, w, F)= \overset{\circ}{f}(r, w, F), \ \lambda(t_0, r) =
\overset{\circ}{\lambda}(r) , \ \mu(t_0, r) =
\overset{\circ}{\mu}(r),
\end{eqnarray*}
and want to show that the corresponding solution exists for all $t
\in ]0, t_0]$. The paper is organized as follows. In section 2 we
first prove that for the cases $\Lambda \leq 0$ and $k \geq 0$ the
solutions obtained in Proposition \ref{p:2.1} below exist on the
whole interval $]0, t_0]$. Next for $\Lambda < 0$ and $k =-1$, and
in the case $\Lambda > 0$, those solutions exist on $]0, t_0]$
provided the initial data are sufficiently small. Later on we
investigate the spatially homogeneous case. In section 3 the
asymptotic behaviour of solutions as $t \to 0$ is investigated for
small data. The last section summarizes all the results which are
known on the contracting direction, for all values of $k$ and
$\Lambda$, including $\Lambda=0$ and both homogeneous and
inhomogeneous models.
\section{On past global existence}
\subsection{The inhomogeneous case}

In this section we make use of the continuation criterion in the
following local existence result:
\begin{proposition} \label{p:2.1}
Let
 $\overset{\circ}{f} \in C^{1}(\mathbb{R}^{2} \times [0, \infty[)$
 with $\overset{\circ}{f}(r+1,w,F) = \overset{\circ}{f}(r,w,F)$
 for $(r,w,F) \in \mathbb{R}^{2} \times [0, \infty[$,
 $\overset{\circ}{f}\geq 0$, and
 \begin{eqnarray*}
 w_{0} := \sup \{ |w| | (r,w,F) \in {\rm supp} \overset{\circ}{f} \} <
 \infty
 \end{eqnarray*}
\begin{eqnarray*}
 F_{0} := \sup \{ F | (r,w,F) \in {\rm supp} \overset{\circ}{f} \} <
 \infty
 \end{eqnarray*}
 Let $\overset{\circ}{\lambda} \in C^{1}(\mathbb{R})$,
 $\overset{\circ}{\mu} \in C^{2}(\mathbb{R})$ with
 $\overset{\circ}{\lambda}(r) = \overset{\circ}{\lambda}(r+1)$,
$\overset{\circ}{\mu}(r) = \overset{\circ}{\mu}(r+1)$ for $r \in
\mathbb{R}$, and
\begin{eqnarray*}
\overset{\circ}{\mu}'(r) =
 -4 \pi t_0 e^{\overset{\circ}{\lambda} +
\overset{\circ}{\mu}}\overset{\circ}{j}(r) = -\frac{4
\pi^{2}}{t_0}e^{\overset{\circ}{\lambda}+ \overset{\circ}{\mu}}
\int_{-\infty}^{\infty} \int_{0}^{\infty}w
\overset{\circ}{f}(r,w,F) dF dw, \ \  r \in \mathbb{R}
\end{eqnarray*}
Then there exists a unique, left maximal, regular solution $(f,
\lambda, \mu)$ of (\ref{eq:1.2})-(\ref{eq:1.6}) with $(f, \lambda,
\mu)(t_0) = (\overset{\circ}{f}, \overset{\circ}{\lambda},
\overset{\circ}{\mu})$ on a time interval $]T, t_0]$ with $T \in
[0, t_0[$. If
\begin{eqnarray*}
\sup \{ |w| | (t, r, w, F) \in {\rm supp} f \} < \infty
\end{eqnarray*}
and
\begin{eqnarray*}
\sup \{ e^{2\mu(t, r)} | r \in \mathbb{R}, t\in ]T, t_0] \} <
\infty
\end{eqnarray*}
then $T = 0$.
\end{proposition}
This is the content of theorems 3.1 and 3.2 in \cite{tchapnda1}.
For a {\it regular solution} all derivatives which appear in the
system exist and are continuous by definition (cf
\cite{tchapnda1}). We prove the following result:
\begin{theorem} \label{t:3.1}
Consider a solution of the Einstein-Vlasov system with $k \geq 0$
and $\Lambda \leq 0$ and initial data given for $t=t_0 > 0$. Then
this solution exists on the whole interval $(0, t_0]$.
\end{theorem}
{\bf Proof} Observe that since there are two choices between two
alternatives, this covers four cases in total namely
\begin{equation*}
(\Lambda,k) \in \{(0,0),(0,1)\}, \  (\Lambda<0,k=0) \ \textrm{or}
\ (\Lambda<0,k=1).
\end{equation*}
In the case $\Lambda=0$, $k=0$ the theorem is a special case of
what was proved by M. Weaver in \cite{weaver}. We then have to
prove the other three cases.

The strategy of the proof is the following : suppose we have a
solution on an interval $(t_1,t_0]$ with $t_1>0$. We want to show
that the solution can be extended to the past. By consideration of
the maximal interval of existence this will prove the assertion.

Firstly let us prove that under the hypotheses of the theorem,
$\mu$ is bounded above.\\ For
\begin{equation}\label{eq:3.1}
\frac{d}{dt}(te^{-2\mu})=-k+\Lambda t^2 -8 \pi t^{2}p \le 0.
\end{equation}
So $te^{-2 \mu}$ cannot increase towards the future, i.e. it
cannot decrease towards the past. Thus on $(t_1, t_0]$, $e^{-2
\mu}$ must remain bounded away from zero and hence $\mu$ is
bounded above.

Recalling that the analogue of Proposition \ref{p:2.1} for
$\Lambda =0$ was proved by G. Rein in \cite{rein2}, we can deduce
that for all three cases being considered it is enough to bound
$w$ to get existence up to $t=0$, using Proposition \ref{p:2.1}.

So let us prove that $w$ is bounded.\\
Consider the following rescaled version of $w$, called $u_1$,
which has been inspired by the work of M. Weaver \cite[p.
1090]{weaver}:
\begin{equation*}
u_1 = \frac{e^{\mu}}{2t}w.
\end{equation*}
If we prove that $\mu$ is bounded below then the boundedness of
$u_1$ will imply the boundedness of $w$. So let us show that $\mu$
is bounded below under the assumption that $u_1$ is bounded.\\
Using the first equality in (\ref{eq:3.1}) and transforming the
integral defining $p$ to $u_1$ as an integration variable instead
of $w$ yields
\[
p= \int_{-\infty}^{\infty}\int_{0}^{\infty}\frac{8\pi t
e^{-3\mu}u_{1}^{2}}{\sqrt{1+4t^{2}e^{-2\mu}u_{1}^{2}+F/t^{2}}}f dF
du_{1},
\]
the integrand can then be estimated by $4\pi e^{-2\mu}|u_1|$.
Thus, using the bound for $u_1$, $p$ can be estimated by $C
e^{-2\mu}$ and so (\ref{eq:3.1}) implies that
\begin{equation*}
|\frac{d}{dt}(te^{-2\mu})|\le C(1+te^{-2\mu}),
\end{equation*}
integrating this with respect to $t$ over $[t,t_0]$ yields
\begin{equation*}
te^{-2\mu}(t,r) \le
t_{0}e^{-2\mu}(t_{0},r)+\int_{t}^{t_0}C\left(1+se^{-2\mu}(s,r)\right)ds,
\end{equation*}
which implies by the Gronwall inequality that $te^{-2\mu}$ is
bounded on $(t_1,t_0]$ that is $\mu$ is bounded below on the given
time interval.

The next step is to prove that $u_1$ is bounded. To this end, it
suffices to get a suitable integral inequality for $\bar u_1$,
where $\bar u_1$ is the maximum modulus of $u_1$ on support of $f$
at a given time. In the vacuum case there is nothing to be proved
and therefore we can assume without loss of generality that
$\bar{u}_1 > 0$.

We can compute $\dot u_1$ :
\begin{equation*}
\dot u_1 = - \frac{e^{\mu}}{2t^2}w+\frac{e^{\mu}}{2t}w(\dot
\mu+\dot r \mu' )+ \frac{e^{\mu}}{2t}\dot w
\end{equation*}
i.e.
\begin{equation}\label{eq:3.2}
\dot u_1 = \left( \dot \mu + \dot r \mu'-\frac{1}{t}\right)u_1 +
\frac{e^{\mu}}{2t}\dot w
\end{equation}
but we have
\begin{equation*}
\mu'= -4 \pi t e^{\mu + \lambda}j, \ \dot r =
\frac{e^{\mu-\lambda}w}{\sqrt{1+w^2+F/t^2}}
\end{equation*}
and
\begin{equation*}
\dot w = 4 \pi t e^{2\mu}(j\sqrt{1+w^2+F/t^2}-\rho w)+\frac{1+k
e^{2\mu}}{2t}w-\frac{\Lambda}{2}t e^{2\mu}w
\end{equation*}
so that multiplying equation (\ref{eq:3.2}) by $2 u_1$ yields the
following :
\begin{equation} \label{eq:3.3}
  \frac{d}{dt} (u_1^2) = 2e^{2\mu}\left[-4 \pi t(\rho-p)+ \frac{k}{t}- \Lambda t\right] u_1
  + 4 \pi e^{3\mu}j \frac{u_1(1+F/t^{2})}{\sqrt{1+4 t^{2} e^{-2\mu} u_{1}^{2}
  + F/t^{2}}}.
\end{equation}
Now the modulus of the first term on the right hand side of
equation (\ref{eq:3.3}) will be estimated. What we need to
estimate is $e^{2\mu}(\rho-p)\bar u_1^2$. For convenience let
$\log_+$ be defined by $\log_+(x)=\log x$ when $\log x$ is
positive and $\log_+(x)=0$, otherwise. Then estimating the
integral defining $\rho-p$ shows that
\begin{equation*}
\rho-p \le C(1+ \log_+(\bar u_1)-\mu).
\end{equation*}
The expression $-\mu$ is not under control ; however the
expression we wish to estimate contains a factor $e^{2\mu}$. The
function $\mu \mapsto -\mu e^{2\mu}$ has an absolute maximum at
$-1/2$ where it has the value $(1/2)e^{-1}$. Thus the first term
on the right hand side of equation (\ref{eq:3.3}) can be estimated
in modulus by $C \bar u_1^2 (1+ \log_+(\bar u_1))$.

Next the modulus of the second term on the right hand side of
equation (\ref{eq:3.3}) will be estimated. Using equation
(\ref{eq:1.9}) defining $j$ it can be estimated by $C \bar w^2$,
i.e.
\begin{equation*}
j \le C \bar u_{1}^{2}e^{-2\mu},
\end{equation*}
so that it suffices to estimate the quantity
\begin{equation}\label{eq:3.4}
\frac{\bar u_{1}^{2}(1+F/t^{2})|u_1|}{\sqrt{1+4 t^{2} e^{-2\mu}
u_{1}^{2}
  + F/t^{2}}}
\end{equation}
in order to estimate the modulus of the second term on the right
hand side of equation (\ref{eq:3.3}). But since $\mu$ and $t^{-1}$
are bounded on the interval being considered, the quantity
(\ref{eq:3.4}) can be estimated by $C\bar u_{1}^{2}$. Thus adding
the estimates for the modulus of the first and second terms on the
right hand side of (\ref{eq:3.3}) allows us to deduce from
(\ref{eq:3.3}), since $\log_+ x = (1/2) \log_+ (x^2)$, that
\begin{equation}\label{eq:3.5}
  |\frac{d}{dt} (u_1^2)| \le C \bar u_1^2 (1+ \log_+(\bar u_1^2)).
\end{equation}
Integrating this in $t$ and using the estimates gives the
following integral inequality for $\bar u_1^2$
\begin{equation}\label{eq:3.6}
  \bar u_1^2(t) \le \bar u_1^2(t_0)+C\int_{t}^{t_0} \bar u_1^2(s) \left(1+ \log_+(\bar
  u_1^2)(s)\right)ds.
\end{equation}
Now let us prove that this integral inequality allows $\bar u_1^2$
and hence $\bar u_1$ to be bounded. The integral inequality
(\ref{eq:3.6}) is of the form
\begin{equation*}
 v(t) \le v(t_0)+C\int_{t}^{t_0} v(s) \left(1+ \log_+(v(s)\right)ds
\end{equation*}
where we have written $v= \bar u_1^2$. By the comparison principle
for solutions of integral equations it is enough to show that the
solution of the integral equation
\begin{equation*}
 v_1(t) = v_1(t_0)+C\int_{t}^{t_0} v_1(s) \left(1+ \log_+(v_1(s)\right)ds
\end{equation*}
is bounded. This solution $v_1(t)$ is a non-increasing function.
Thus either $v_1(t) \leq e$ everywhere, in which case the desired
conclusion is immediate, or there is some $t_2$ in $(t_1, t_0]$
such that $v_1(t)\geq e$ on $(t_1, t_2]$, we take $t_2$ maximal
with that property. In this second case it follows that on the
interval $(t_1, t_2]$ the inequality
\begin{equation*}
 v_1(t) \leq C\left(1+ \int_{t}^{t_2} v_1(s) \log v_1(s)ds\right)
\end{equation*}
holds for a constant $C$. The boundedness of $v_1(t)$ follows from
that of the solutions of the differential equation $\dot v_2(t) =
C v_2(t) \log v_2(t)$. In fact we get a bound like exp(exp $t$)
for $v_1(t)$. Either case $v_1(t)$ is bounded. Thus we conclude
that $\bar u_1^2$ and hence $u_1$ is bounded i.e. $w$ is bounded
and the proof of the theorem is complete.$\Box$

It is important to note that in the case $(\Lambda=0, k=1)$ the
result proved in Theorem \ref{t:3.1} is new and so strengthens the
existence up to $t=0$ for small data obtained in \cite{rein2}.

Next we have the following result which generalizes Theorem 4.1 in
\cite{rein2} to the case with non-zero cosmological constant
$\Lambda$. Since there are only minor changes in the proof, we
omit it here.
\begin{theorem} \label{t:3.2}
Let $(\overset{\circ}{f}, \overset{\circ}{\lambda},
\overset{\circ}{\mu})$ be initial data as in Proposition
\ref{p:2.1}, and assume that $e^{-2\overset{\circ}{\mu}(r)}-
\frac{4}{3}\Lambda t_{0}^{2}-2 > 0$ for $r \in \mathbb{R}$ and
$c>0$ with
\begin{equation*}
c:= \frac{1}{2}(1-\parallel e^{2\overset{\circ}{\mu}}\parallel)-
10 \pi^2 w_0 F_0
\sqrt{1+w_{0}^{2}+F_{0}/t_{0}^{2}}\parallel\overset{\circ}{f}\parallel
\frac{\parallel e^{2\overset{\circ}{\mu}}\parallel}{1-\parallel
e^{2\overset{\circ}{\mu}}\parallel} \ \textrm{if} \ k=-1 \
\textrm{and} \ \Lambda < 0 ,
\end{equation*}
and for $\Lambda>0$
\begin{displaymath}
c:= \left\{ \begin{array}{ll} \frac{1}{2} \left( 1-\frac{\Lambda
t_{0}^{2}\parallel e^{2\overset{\circ}{\mu}}\parallel}{1-
\frac{\Lambda}{3}t_{0}^{2}\parallel
e^{2\overset{\circ}{\mu}}\parallel } \right)- 10 \pi^2 w_0 F_0
\sqrt{1+w_{0}^{2}+F_{0}/t_{0}^{2}}\parallel\overset{\circ}{f}\parallel
\frac{\parallel e^{2\overset{\circ}{\mu}}\parallel}{1-
\frac{\Lambda}{3}t_{0}^{2}\parallel
e^{2\overset{\circ}{\mu}}\parallel},\\
 \textrm{if $k=0$ or k=1} \\
\frac{1}{2}\left( 1-\frac{(\Lambda t_{0}^{2}+1)\parallel
e^{2\overset{\circ}{\mu}}\parallel}{1-
(\frac{\Lambda}{3}t_{0}^{2}+1)\parallel
e^{2\overset{\circ}{\mu}}\parallel}\right)- 10 \pi^2 w_0 F_0
\sqrt{1+w_{0}^{2}+F_{0}/t_{0}^{2}}\parallel\overset{\circ}{f}\parallel
\frac{\parallel e^{2\overset{\circ}{\mu}}\parallel}{1-
(\frac{\Lambda}{3}t_{0}^{2}+1)\parallel
e^{2\overset{\circ}{\mu}}\parallel},\\ \textrm{if $k=-1$}.
\end{array} \right.
 \end{displaymath}
Then the corresponding solution exists on the interval $]0, t_0]$,
and
\begin{equation*}
|w| \le w_0 t^c, \ (r,w,F) \in \supp f(t), \ t \in ]0, t_0].
\end{equation*}
\end{theorem}

\begin{remark}\label{r:2.4} Note that in the case $\Lambda <0$ and $k \geq 0$ it
suffices to let
\begin{equation*}
 c := \frac{1}{2}-
10 \pi^2 w_0 F_0
\sqrt{1+w_{0}^{2}+F_{0}/t_{0}^{2}}\parallel\overset{\circ}{f}\parallel
\parallel
e^{2\overset{\circ}{\mu}}\parallel
\end{equation*}
 in the hypotheses of Theorem \ref{t:3.2} in order to obtain the
 same conclusion in the latter theorem.
 \end{remark}

It may be asked what happens for general data. In order to answer
this question we begin here by examining the spatially homogeneous
case. Note that for $\Lambda=0$ more information is available in
the homogeneous case in \cite{rendall2}.
\subsection{The spatially homogeneous case}
In this subsection we want to prove that spatially homogeneous
solutions (i.e. solutions which are independent of $r$) exist on
the whole interval $(0, t_0]$ for general initial data in the case
of hyperbolic symmetry and negative $\Lambda$ as well as in the
positive $\Lambda$ case.

To this end we use the continuation criterion stated in
Proposition \ref{p:2.1}. Since the proof for the boundedness of
$w$ in Theorem \ref{t:3.1} depends neither on the sign of $k$ nor
that of $\Lambda$, the only thing to do here is to bound $\mu$ in
order to get existence up to $t=0$. We prove in three
steps that $\mu$ is bounded above.\\
{\it Step 1} As a first step we use the notations and the method
of the proof for Lemma 3 in \cite{weaver} to obtain a 'lower
bound' of the quantity
\begin{equation*}
\int_{\mathbb{R}^{3}}f \frac{v_1^2}{|v_0|} dv_1 dv_2 dv_3.
\end{equation*}
By equation (4) in \cite{weaver},
\begin{equation*}
v_0 = - \sqrt{\alpha e^{2\nu-2U}+\alpha v_1^2+\alpha e^{2\nu-4U}
v_2^2+\alpha \tilde{t}^{-2}e^{2 \nu}(v_3-A v_2)^{2}},
\end{equation*}
in order to save some notation we have denoted by $\tilde{t}$ the
area of the symmetry orbits in \cite{weaver}. Taking $A=0$ we
obtain the following inequality, since $|v_2|\leq \bar{v_2}$ and
$|v_3|\leq \bar{v_3}$ :
\begin{equation*}
|v_0| \leq \sqrt{K+ \alpha v_1^2},
\end{equation*}
with
\begin{equation*}
K =\alpha e^{2\nu-2U}+\alpha e^{2\nu-4U} \bar{v_2}^{2}+\alpha
\tilde{t}^{-2}e^{2 \nu}\bar{v_3}^{2},
\end{equation*}
so that
\begin{equation*}
\frac{v_1^2}{|v_0|} \geq \frac{v_1^2}{\sqrt{K+\alpha v_1^2}}.
\end{equation*}
If $|v_1|> \delta$ then it follows that
\begin{equation*}
\frac{v_1^2}{\sqrt{K+\alpha v_1^2}} \geq
\frac{\delta^2}{\sqrt{K+\alpha \delta ^2}}.
\end{equation*}
Now the argument of the proof for Lemma 3 in \cite{weaver} applies
here and the only thing we need to change there is that we replace
the displayed equation leading to equation (10) by what follows :
\begin{align}\label{eq:1}
 \int_{\mathbb{R}^{3}}f \frac{v_1^2}{|v_0|}
dv_1 dv_2 dv_3 & = \int_{\mathbb{R}^{2}}\int_{-\delta}^{\delta}f
\frac{v_1^2}{|v_0|} dv_1 dv_2 dv_3 +
\int_{\mathbb{R}^{2}}\int_{|v_1| > \delta}f \frac{v_1^2}{|v_0|}
dv_1 dv_2 dv_3
\nonumber\\
& \geq \frac{\delta^2}{\sqrt{K+\alpha \delta
^2}}\int_{\mathbb{R}^{2}}\left(\int_{|v_1| > \delta}f
 dv_1 \right) dv_2 dv_3
\nonumber\\
 & \geq \frac{\delta^2 b}{\sqrt{K+\alpha \delta
^2}}
\nonumber\\
& \geq \frac{C}{\sqrt{K+\alpha}},
\end{align}
where $C$ is a positive constant.\\
{\it Step 2} Now we translate the latter inequality into our
familiar notation of plane symmetry. We have\\
$v_0=-e^{2\mu}p^0/2t$, $v_1=e^{2\lambda}p^1$, $v_2=t^2 p^2$,
$v_3=t^2 p^3$, $\tilde{t}=t^2$, $U= \log t$, $\nu = \lambda +\log
t$ and $\alpha=
(1/4)t^{-2}e^{2(\mu - \lambda)}$.\\
Then
\begin{equation*}
K+ \alpha = (1/4)t^{-4}e^{2 \mu}[t^2 e^{-2
\lambda}+t^2+\bar{v_2}^{2}+\bar{v_3}^{2}],
\end{equation*}
so that
\begin{equation}\label{eq:2}
  \frac{1}{\sqrt{K+\alpha}} \geq \frac{C t^2
  e^{\lambda-\mu}}{e^{2\lambda}+1}.
\end{equation}
On the other hand the jacobian determinant of the transformation
$(p^1, p^2, p^3)\mapsto (v_1, v_2, v_3)$ gives
\begin{equation*}
\det \left(\frac{\partial(v_1, v_2, v_3)}{\partial(p^1, p^2,
p^3)}\right)=t^4 e^{2 \lambda};
\end{equation*}
it follows that
\begin{equation*}
\int_{\mathbb{R}^{3}}f \frac{v_1^2}{|v_0|} dv_1 dv_2 dv_3 = 2
\int_{\mathbb{R}^{3}}f t^5 e^{6\lambda-2\mu}\frac{(p^1)^2}{|p^0|}
dp^1 dp^2 dp^3;
\end{equation*}
but we have $w:= e^\lambda p^1$, $F:= t^4[(p^2)^2+(p^3)^2]$ so
that
\begin{equation*}
|\det \left(\frac{\partial(w, F, p^3)}{\partial(p^1, p^2,
p^3)}\right)|=2t^4 e^{\lambda}\sqrt{Ft^{-4}-(p^3)^2};
\end{equation*}
therefore calculating the latter integral implies that
\begin{equation}\label{eq:3}
\int_{\mathbb{R}^{3}}f \frac{v_1^2}{|v_0|} dv_1 dv_2 dv_3 =
(1/2)t^3 e^{3 \lambda-\mu}p.
\end{equation}
Thus using equations (\ref{eq:1}), (\ref{eq:2}) and (\ref{eq:3})
yields the following estimate :
\begin{equation}\label{eq:4}
p \geq \frac{C t^{-1}e^{-2\lambda}}{e^{2 \lambda}+1},
\end{equation}
{\it Step 3} Boundedness for $\mu$.\\
Now we have
\begin{equation*}
2te^{-2\mu}(\dot{\lambda}+\dot{\mu})= 8 \pi t^2(\rho+p) \geq 0
\end{equation*}
which implies that $\lambda+\mu$ is increasing and so in the past
time direction the estimate $e^{\lambda+\mu} \leq C$ i.e. $e^{-2
\lambda} \geq C e^{2 \mu}$ holds. Equation (\ref{eq:4}) then
implies
\begin{equation}\label{eq:5}
p \geq C\frac{e^{2\mu}}{e^{-2 \mu}+1}.
\end{equation}
The estimate (\ref{eq:5}) shows that if $\mu$ is arbitrarily
large, $p$ will become also large and will dominate the other
terms of the right hand side in the equality
\begin{equation*}
\frac{d}{dt}(te^{-2\mu})=-k+\Lambda t^2-8 \pi t^2 p
\end{equation*}
and then $\frac{d}{dt}(te^{-2\mu})$ will become negative, i.e.
there is a constant $L$ such that the following implication holds
\begin{equation*}
\mu \geq L \ \ \Rightarrow \ \ \frac{d}{dt}(te^{-2\mu})<0.
\end{equation*}
Now fix $t_1$ in any interval $(t_*, t_0]$. Then either $\mu(t_1)
\leq L$ or $\mu(t_1) > L$. In this second case define
\begin{equation*}
t_2 := \sup \left\{t \in (t_*, t_0]: \mu \geq L \ {\rm on} \ [t_1,
t]\right\},
\end{equation*}
we thus have either $t_2 < t_0$ or $t_2=t_0$. Since $\mu(t)\geq L$
for all $t \in [t_1, t_2]$, it follows that
\begin{equation*}
t_1 e^{-2 \mu(t_1)} \geq t_2 e^{-2 \mu(t_2)}
\end{equation*}
that is
\begin{align*}
 e^{-2 \mu(t_1)} & \leq  e^{-2 \mu(t_2)}(t_1/t_2)\\
 & \leq
 \begin{cases}
 e^{2L}(t_1/t_2) \ {\rm if} \ t_2 < t_1\\
e^{2\mu(t_0)}(t_1/t_0) \ {\rm if} \ t_2 = t_0
\end{cases}
\end{align*}
either case, $e^{2 \mu}$ and so $\mu$ is uniformly bounded in $t$
in any interval $(t_*, t_0]$.

We have proven the following
\begin{theorem}
Consider a spatially homogeneous solution of the Einstein-Vlasov
system with $\Lambda > 0$ or $(\Lambda < 0, k=-1)$, and initial
data given for $t=t_0 > 0$. Then this solution exists on the whole
interval $(0, t_0]$.
\end{theorem}
\section{On past asymptotic behaviour}
In this section we examine the behaviour of solutions as $t \to
0$.
\subsection{The initial singularity} First we analyze the
curvature invariant
$R_{\alpha\beta\gamma\delta}R^{\alpha\beta\gamma\delta}$ called
the Kretschmann scalar in order to prove that there is a spacetime
singularity
\begin{theorem}\label{t:3.3}
Let $(f,\lambda,\mu)$ be a regular solution of the
surface-symmetric Einstein-Vlasov system with cosmological
constant on the interval $]0,t_0]$ with small initial data as
described in Theorem \ref{t:3.2} and in Remark \ref{r:2.4}. Then
\begin{displaymath}
\lim_{t \to
0}(R_{\alpha\beta\gamma\delta}R^{\alpha\beta\gamma\delta})(t,r) =
\infty,
\end{displaymath}
uniformly in $r \in \mathbb{R}$.
\end{theorem}
{\bf Proof} We can compute the Kretschmann scalar as in
\cite{rein2} and obtain
\begin{align}
R_{\alpha\beta\gamma\delta}R^{\alpha\beta\gamma\delta} & =
4[e^{-2\lambda}(\mu''+\mu'(\mu'-\lambda'))-e^{-2\mu}(\ddot
\lambda+ \dot \lambda(\dot \lambda- \dot \mu))]^{2} \nonumber \\
& + \frac{8}{t^2}[e^{-4\mu} \dot \lambda^{2}+e^{-4\mu} \dot
\mu^{2}-2e^{-2(\lambda+\mu)}(\mu')^{2}] \nonumber \\
& +\frac{4}{t^4}(e^{-2\mu}+k)^{2} \nonumber \\
& =: K_{1}+K_{2}+K_{3}
\end{align}
Since $K_1$ is nonnegative it can be dropped.\\ Now let us
distinguish the cases $\Lambda >0$ and $\Lambda <0$.

Case $\Lambda >0$ :\\ In this case we use the same argument as in
\cite{rein2} to estimate $K_2$. We then obtain
\begin{align}\label{eq:3.16}
K_2 & \ge  \frac{8}{t^2}\left[\left(4\pi
t(\rho-p)\frac{k+e^{-2\mu}}{2t}\right)^{2}+\left(
\frac{k+e^{-2\mu}}{2t}- \Lambda t
\right)^{2}-\frac{\Lambda^2}{2}t^2+4\pi t^2 \Lambda (\rho-p)
\right] \\
& \ge -4 \Lambda^{2} \nonumber
\end{align}
since $4\pi t^2 \Lambda (\rho-p) \ge 0$.\\ Recalling the
expression for $e^{-2\mu}$ we get
\begin{align} \label{eq:3.17}
e^{-2 \mu}+k & = \frac{t_{0}(e^{-2 \overset{\circ}{\mu}(r)} +
k)}{t} + \frac{8 \pi}{t}\int_{t}^{t_{0}}s^{2}p(s, r) ds +
\frac{\Lambda}{3t}(t^{3} - t_{0}^{3}) \nonumber\\
& \ge \frac{t_0(\inf e^{-2
\overset{\circ}{\mu}}+k-\frac{\Lambda}{3}t_{0}^{2})}{t}
\end{align}
thus
\begin{equation*}
K_3 = \frac{4}{t^4}(e^{-2\mu}+k)^{2} \ge \frac{4
t_{0}^{2}}{t^6}\left(\inf e^{-2
\overset{\circ}{\mu}}+k-\frac{\Lambda}{3}t_{0}^{2}\right)^{2}
\end{equation*}
and so
\begin{equation*}
 (R_{\alpha\beta\gamma\delta}R^{\alpha\beta\gamma\delta})(t,r) \ge
 \frac{4
t_{0}^{2}}{t^6}\left(\inf e^{-2
\overset{\circ}{\mu}}+k-\frac{\Lambda}{3}t_{0}^{2}\right)^{2} -4
\Lambda^2, \ t \in ]0,t_0], \ r \in \mathbb{R}, \Lambda >0,
\end{equation*}
and the assertion is proved for $\Lambda>0$.

Case $\Lambda < 0$\\
We have by (\ref{eq:3.16}) :
\begin{equation*}
K_2 \ge -4\Lambda^2 +32\pi \Lambda (\rho-p).
\end{equation*}
Now we use the estimate for $w$ in Theorem \ref{t:3.2} so that
\begin{equation}\label{eq:3.18}
(\rho-p)(t,r) \le \rho(t,r) =
\frac{\pi}{t^2}\int_{-P(t)}^{P(t)}\int_{0}^{F_0}\sqrt{1+w^2+F/t^2}f(t,r,w,F)dF
dw \le C t^{-3+c},
\end{equation}
where $c$ is defined in Theorem \ref{t:3.2} and in Remark \ref{r:2.4}.\\
Thus
\begin{equation*}
K_2 \ge -4\Lambda^2 +C \Lambda t^{-3+c}.
\end{equation*}
(\ref{eq:3.17}) becomes in this case ($\Lambda < 0$)
\begin{align*}
e^{-2 \mu}+k  \ge \frac{t_0(\inf e^{-2
\overset{\circ}{\mu}}+k)}{t},
\end{align*}
therefore
\begin{equation*}
K_3 \ge \frac{4 t_{0}^{2}}{t^6}\left(\inf e^{-2
\overset{\circ}{\mu}}+k\right)^{2}
\end{equation*}
so that
\begin{eqnarray*}
 (R_{\alpha\beta\gamma\delta}R^{\alpha\beta\gamma\delta})(t,r) \ge
 \frac{4
t_{0}^{2}}{t^6}\left(\inf e^{-2 \overset{\circ}{\mu}}+k
\right)^{2} +C \Lambda t^{-3+c}-4 \Lambda^2, \ t \in ]0,t_0], \ r
\in \mathbb{R},
\end{eqnarray*}
that is the assertion in the theorem holds for $\Lambda < 0$ as
well, and the proof is complete. $\Box$

Next we prove that the singularity at $t=0$ is a crushing
singularity i.e. the mean curvature of the surfaces of constant
$t$ (cf. \cite[(1.0.2)]{christodoulou} for a definition) blows up,
also it is a velocity dominated singularity i.e the generalized
Kasner exponents have limits as $t \to 0$. We have the same
results as in the vanishing cosmological constant case
\cite{rein2}, the argument of the proof in that case also applies
here.
\begin{theorem}\label{t:3.4}
Let $(f,\lambda,\mu)$ be a solution of the surface-symmetric
Einstein-Vlasov system with cosmological constant on the interval
$]0,t_0]$, assume that the initial data satisfy $e^{-2
\overset{\circ}{\mu}(r)}-\frac{\Lambda}{3}t_{0}^{2}-1>0$ for $r
\in \mathbb{R}$. Let
\begin{equation*}
K(t,r):=-e^{-\mu}\left(\dot \lambda(t,r)+\frac{2}{t}\right)
\end{equation*}
which is the mean curvature of the surfaces of constant $t$.
 Then
\begin{displaymath}
\lim_{t \to 0}K(t,r) = -\infty,
\end{displaymath}
uniformly in $r \in \mathbb{R}$.
\end{theorem}
{\bf Proof} In fact we use the same argument as in \cite{rein2}
and obtain the following inequalities :\\
if $\Lambda >0$ then
\begin{align*}
K(t,r) \le - \frac{t_{0}^{1/2}(\inf e^{-2
\overset{\circ}{\mu}}+k-\frac{\Lambda}{3}t_{0}^{2})^{1/2}}{t^{3/2}}
\ \textrm{for $k=0$ or $k=-1$},
\end{align*}
and
\begin{align*}
K(t,r) \le - \frac{t_{0}^{1/2}(\inf e^{-2
\overset{\circ}{\mu}}-\frac{\Lambda}{3}t_{0}^{2})^{1/2}}{t^{3/2}}
\ \textrm{for $k=1$};
\end{align*}
whereas if $\Lambda<0$,
\begin{align*}
K(t,r) \le -\frac{3}{2} \frac{t_{0}^{1/2}(\inf e^{-2
\overset{\circ}{\mu}}+k)^{1/2}}{t^{3/2}} -
\frac{\Lambda}{2}\frac{t_{0}^{-1/2}(\inf e^{-2
\overset{\circ}{\mu}}+k)^{-1/2}}{t^{-3/2}} \ \textrm{for $k=0$ or
$k=-1$},
\end{align*}
and
\begin{align*}
K(t,r) \le -\frac{e^{-\mu}}{t}-\frac{\Lambda}{2}t e^{\mu} \le -
\frac{t_{0}^{1/2}(\inf e^{-
\overset{\circ}{\mu}})}{t^{3/2}}-\frac{\Lambda}{2}\frac{t_{0}^{-1/2}(\inf
e^{- \overset{\circ}{\mu}})^{-1}}{t^{-3/2}} \ \textrm{for $k=1$}.
\ \Box
\end{align*}
\begin{theorem}\label{t:3.5}
Let $(f,\lambda,\mu)$ be a regular solution of the
surface-symmetric Einstein-Vlasov system with cosmological
constant on the interval $]0,t_0]$ with small initial data as
described in Theorem \ref{t:3.2} and in Remark \ref{r:2.4}. Then
\begin{displaymath}
\lim_{t\to 0} \frac{K_{1}^{1}(t,r)}{K(t,r)} = -\frac{1}{3} ; \
\lim_{t\to 0} \frac{K_{2}^{2}(t,r)}{K(t,r)} = \lim_{t\to 0}
\frac{K_{3}^{3}(t,r)}{K(t,r)} = \frac{2}{3},
\end{displaymath}
uniformly in $r \in \mathbb{R}$,\\
where
\begin{equation*}
\frac{K_{1}^{1}(t,r)}{K(t,r)}, \ \frac{K_{2}^{2}(t,r)}{K(t,r)}, \
\frac{K_{3}^{3}(t,r)}{K(t,r)}
\end{equation*}
are the generalized Kasner exponents.
\end{theorem}
Note that in this theorem some smallness assumption on the initial
data is necessary (cf. \cite[p.148]{rein1}).
\subsection{Determination of the leading asymptotic behaviour} In
this subsection we determine the explicit leading behaviour of
$\lambda$, $\mu$, $\dot{\lambda}$, $\dot{\mu}$, $\mu'$ in the case
of small data.

We have
\begin{equation*}
  \frac{d}{dt}(t e^{-2\mu})= \Lambda t^{2}-k-8\pi t^2 p.
\end{equation*}
By Theorem \ref{t:3.2} and Remark \ref{r:2.4},
\begin{eqnarray}\label{eq:3.21}
 |w| \le Ct^c.
\end{eqnarray}
Using the expression for $p$, we have
\begin{eqnarray*}
  p \le Ct^{-2+3c}
\end{eqnarray*}
so that
\begin{equation*}
  8\pi t^2 p \le C,
\end{equation*}
thus
\begin{equation*}
  \mid \frac{d}{dt}(t e^{-2\mu})\mid \le C,
\end{equation*}
integrating this over $[t_1,t_2]$ yields
\begin{equation}\label{eq:3.22}
 \mid t_{2} e^{-2\mu(t_2)}-t_{1} e^{-2\mu(t_1)}\mid \le C(t_2 -t_1)
\end{equation}
so $s \mapsto s e^{-2\mu(s)}$ verifies the Lipschitz condition
with Lipschitz constant $C$. Thus
\begin{equation*}
t e^{-2\mu(t)} \to L \ \textrm{as $t \to 0$} ;
\end{equation*}
note that $L>0$, using the lower bound on $e^{-2\mu(t)}$.
(\ref{eq:3.22}) then implies
\begin{equation*}
\mid t e^{-2\mu(t)}- L \mid \le C t
\end{equation*}
and so
\begin{equation*}
e^{-2\mu(t)}= \frac{L}{t} +O(1)= \frac{L}{t}(1+O(t))
\end{equation*}
thus
\begin{equation}\label{eq:3.23}
e^{2\mu(t)}= L^{-1}t(1+O(t))
\end{equation}
that is
\begin{equation}\label{eq:3.24}
\mu= \frac{1}{2}\ln t+O(1).
\end{equation}
Now we have
\begin{equation}\label{eq:3.25}
  \dot \lambda = \frac{1}{2}(\Lambda t+8\pi t \rho)e^{2\mu}
  -\frac{1+ke^{2\mu}}{2t}.
\end{equation}
Using (\ref{eq:3.21}) and the expression for $\rho$ we can see
that
\begin{eqnarray*}
 8\pi t \rho \le Ct^{-2+c}
\end{eqnarray*}
thus (\ref{eq:3.25}) implies that
\begin{align*}
 \dot \lambda = \frac{1}{2}[L^{-1}\Lambda
 t^{2}+O(t^{3})+O(t^{-1+c})+O(t^{c})]-\frac{1}{2t}-\frac{k
 L^{-1}}{2}+O(t)
\end{align*}
that is, using the fact that $-1+c <0$,
\begin{eqnarray}\label{eq:3.26}
 \dot \lambda =  -\frac{1}{2t}+O(t^{-1+c})
\end{eqnarray}
and so
\begin{eqnarray}\label{eq:3.27}
\lambda =  -\frac{1}{2}\ln t+O(t^{c}).
\end{eqnarray}
Now using (\ref{eq:3.21}) and the expression for $j$ we can see
that
\begin{eqnarray*}
  j \le Ct^{-2+c}
\end{eqnarray*}
and thus using equation $\mu' = -4\pi t e^{\lambda+\mu}j$ we
obtain
\begin{equation*}
 \mu' = -4\pi t L^{-1/2}t^{1/2}[1+O(t)]\left[t^{-1/2}(1+O(t^c))\right]O(t^{-2+c})
\end{equation*}
i.e.
\begin{eqnarray}\label{eq:3.28}
  \mu' = O(t^{-1+c});
\end{eqnarray}
we have used equations (\ref{eq:3.23}) and (\ref{eq:3.27}).\\
Recalling that
\begin{equation*}
  \dot \mu = \frac{1}{2}(-\Lambda t+8\pi t p)e^{2\mu}
  +\frac{1+ke^{2\mu}}{2t},
\end{equation*}
we use (\ref{eq:3.23}) and the fact that $8\pi t p \le C$ to
obtain
\begin{equation}\label{eq:3.29}
\dot \mu= \frac{1}{2t}+O(1).
\end{equation}
Thus we have proven the following
\begin{theorem}\label{t:3.6}
Let $(f,\lambda,\mu)$ be a solution of the surface-symmetric
Einstein-Vlasov system with cosmological constant on the interval
$]0,t_0]$ with small initial data as described in Theorem
\ref{t:3.2} and in Remark \ref{r:2.4} for $\Lambda \neq 0$, and in
\cite[Theorem 4.1]{rein2} for $\Lambda =0$. Then the following
properties hold at early times : (\ref{eq:3.24}), (\ref{eq:3.26}),
(\ref{eq:3.27}), (\ref{eq:3.28}), (\ref{eq:3.29}).
\end{theorem}
This theorem shows that the model for the dynamics of the class of
solutions considered here is the Kasner solution (see \cite{wald})
with Kasner exponents $(2/3,2/3,-1/3)$ for which
$\lambda=-\frac{1}{2}\ln t$ and $\mu=\frac{1}{2}\ln t$.
\section{Concluding remarks}
In the contracting direction we have shown the global existence in
the case $(\Lambda \leq 0, k \geq 0)$ for generic initial data.
Similar results have been obtained for small data in the cases
$(\Lambda < 0, k =-1)$ and $\Lambda > 0$. Detailed early-time
asymptotics have been obtained for small data as well. It would be
interesting to examine what happens if the smallness assumption on
data is dropped. In this case the asymptotics need not be
Kasner-like (cf. \cite[p.146]{rein1}).

In the homogeneous case we have proven existence up to $t=0$ for
generic data. Together with the results obtained in
\cite{tchapnda2} this would prove strong cosmic censorship
(cf.\cite{eardley}) in the class of Bianchi I and III solutions of
the Einstein-Vlasov system with positive cosmological constant, if
we obtained curvature blow-up for generic data. In fact proving
cosmic censorship requires proving inextendibility of the maximal
Cauchy development in the future and in the past. Still in the
homogeneous case more information about asymptotics for $\Lambda
=0$ is available in \cite{rendall2}. In the case with $\Lambda$,
such results are not available.

\section*{Acknowledgments} The author
thanks A.D. Rendall for helpful comments. He also thanks N.
Noutchegueme. Support by a research grant from the
VolkswagenStiftung in Federal Republic of Germany is acknowledged.
This work was completed at the Abdus Salam International Centre
for Theoretical Physics in Trieste during a visit under the junior
associateship scheme sponsored by the Swedish International
Development Cooperation Agency (SIDA).

\end{document}